\documentclass[preprintnumbers,amsmath,amssymb,aps]{revtex4}
\usepackage{graphicx}

\newcommand{\bee}{\begin{equation}}
\newcommand{\ee}{\end{equation}}
\newcommand{\beea}{\begin{eqnarray}}
\newcommand{\eea}{\end{eqnarray}}
\newcommand{\gfive}{\gamma_5}
\newcommand{\sign}{{\rm sign}}
\def\Tr{{\rm Tr}}
\newcommand{\imu}{{\rm i}}

\begin{document}

\title{Physics issues in simulations with  dynamical overlap fermions }

\author{Thomas DeGrand}
\author{Stefan Schaefer}
\affiliation{
Department of Physics, University of Colorado,
        Boulder, CO 80309 USA}

\begin{abstract}
We discuss the impact of various improvements on 
simulations of dynamical overlap fermions using the 
Hybrid Monte Carlo algorithm. We focus on the usage of fat links and 
multiple pseudo fermion fields.
\end{abstract}
\maketitle

\section{Introduction}
As lattice simulations of QCD push to ever smaller quark masses, they increasingly benefit
from discretizations which respect chiral symmetry. Simulations with $N_f$ flavors of degenerate overlap
\cite{Neuberger:1997fp,Neuberger:1998my} fermions encode an exact $SU(N_f)\otimes SU(N_f)$ chiral symmetry,
and are the cleanest theoretical description of lattice fermions known to date.
They have several
theoretical advantages as compared to the traditional formulations
using Wilson or staggered quarks. Their correct implementation
of chiral symmetry protects them from exceptional configurations
which plague simulations with Wilson fermions.
Staggered fermions have a $U(1)\otimes U(1)$ chiral symmetry, and are probably the
 easiest fermions to simulate,
but one staggered fermion flavor  corresponds to four  continuum ``tastes.''
It is unknown whether the
standard trick of performing staggered simulations by taking the quarter root of
the fermion determinant is justifiable. Staggered fermions also have a 
complicated low energy effective field theory due to taste-breaking interactions, while the
 low energy dynamics of overlap fermions is basically identical to that of the continuum.
Domain wall fermions involve simulations on a five-dimensional background. Flavor symmetry is unbroken,
 and chiral
symmetry becomes exact as the length of the fifth dimension goes to infinity. Recent large scale simulations
with domain wall fermions claim to have good chiral behavior\cite{Aoki:2004ht}.
 However, a paper\cite{Golterman:2004mf} which appeared nearly
simultaneously, describing requirements for good chiral behavior for matrix elements appropriate to
CP violation in kaons, set somewhat more stringent requirements than the simulation claimed.
We prefer to avoid these questions by working  from the beginning
with exact chiral symmetry for dynamical fermions.

Unfortunately, the much higher cost of applying the Dirac 
operator on a vector  prevents us from using overlap fermions in large scale
simulations. Until now only exploratory studies have been published 
\cite{Bode:1999dd,Fodor:2003bh,Cundy:2004xf}. This paper is also only  exploratory. Its
goal is to investigate 
several methods to improve the performance of the Hybrid Monte Carlo
(HMC) algorithm \cite{Gottlieb:1987mq} and the extension proposed in
\cite{Fodor:2003bh} for these fermions. The major new ingredient
is the use of a differentiable fat link, the ``stout link'' of Peardon
and Morningstar\cite{Morningstar:2003gk},
 as the gauge connection for the fermions. Fat links are well known to 
reduce the computational cost of overlap fermions \cite{DeGrand:2000tf,DeGrand:2002vu}.
 Their bottleneck has been that the
existing forms could not be used in the standard dynamical fermion updating algorithm,
 Hybrid Monte Carlo. Stout links overcome that difficulty and give a
 speedup of about an order of magnitude in computer time compared to simulations
 of overlap fermions with thin links.

The eigenvalues of the overlap Dirac operator lie on a circle in the complex plane.
The overlap operator maps the eigenvalues of a ``kernel'' action onto this circle.
The eigenmodes of these free kernel actions typically lie on arcs, and the eigenvalues
of the kernel in an arbitrary background gauge field can be dense everywhere.
Fat links tend to force the eigenvalues back onto the free-field arcs, reducing the density of eigenmodes
near the center of the overlap circle. Golterman and Shamir\cite{Golterman:2003qe} have recently
suggested that chiral symmetry for  overlap action could be realized differently
 if the eigenmodes of the kernel
whose eigenvalues lie near the
center of the circle  become delocalized in space.
 By reducing the density of eigenmodes, we reduce the probability that nearly
degenerate modes exist and can mix.
We will show that  fat link kernel eigenmodes are more localized than
thin link kernel eigenmodes.

The major difference between dynamical simulations of overlap fermions and other kinds 
  is the discontinuity in the overlap fermionic
action associated with changing the topological sector. 
If carefully treated, away from these topological boundaries, the overlap fermion
 force in Hybrid Monte Carlo
is never singular, and so a larger integration step size can be tolerated.
For us, the force due to the gauge part of the action is actually larger
 than the fermion force.
This part of the force is inexpensive to compute, and it can be easily smoothed by
 using the Sexton-Weingarten\cite{Sexton:1992nu}
multi-scale time-step update.

At a topological boundary, the step in the fermionic action
can be treated by
the algorithm proposed in \cite{Fodor:2003bh} which reflects or refracts
the momentum at the discontinuity in the fermionic 
action in  analogy to classical mechanics. Pseudo-fermions give only a noisy estimator
of the fermion action. The noise suppresses topological changes. This problem can be
partially ameliorated with the multi-pseudofermion  method of
 Hasenbusch\cite{Hasenbusch:2001ne,Hasenbusch:2002ai}.

The use of improved actions is accompanied by the necessity of choosing values
for a large parameter set -- nearly all of which are related to coefficients
of (formally) irrelevant operators. This is only psychologically different from the use of more
standard actions, where essentially all these parameters are set to zero value and ignored.
Generally, there is a wide latitude in the choice of these parameters. As we proceed, we will discuss
our choices (summarizing everything in the conclusion), with a description of what will happen
when the parameters are varied, when it is available. We cannot imagine that anyone
would want to do simulations with precisely our choice of parameters,
 but we believe that any reasonably similar action would produce similar results.

\section{Action and Algorithm}
\subsection{Definitions}
To fix the conventions, let us recall the definition of the overlap Dirac operator $D_{\rm ov}(m_q)$.
It is constructed from some lattice Dirac operator
(in the following called the kernel operator)
 $d(\mu)$ and its associated Hermitian Dirac operator $h(\mu)=\gamma_5 d(\mu)$
with mass term $\mu$. The
 massless Hermitian overlap operator $H(0)=\gfive D_{\rm ov}(0)$ is defined as
\bee
H(0)=R_0\left[\gfive + \epsilon(h(-R_0))\right] \label{eq:hov0}
\ee
with $R_0>0$ the radius of the Ginsparg--Wilson\cite{Ginsparg:1981bj} circle. The matrix sign function 
$\epsilon(h)$ can be defined by 
\bee
\epsilon(h) = \frac{h}{\sqrt{h^2}}=\sum_\lambda \sign(\lambda) 
| \lambda \rangle \langle \lambda | \label{eq:sign}
\ee
with $\lambda$ the eigenvalues and $ |\lambda\rangle$ the eigenvectors of $h(-R_0)$.
 (Whenever no
argument of $h$ is given, $h(-R_0)$ is implied.) The massive overlap Dirac operator is
\bee
D(m) = (1 -\frac{m}{2R_0})D(0)+m,
\ee
and  the
 square of the massive Hermitian overlap operator $H(m)$ can then be written as
\bee
H(m)^2= (1-\frac{m^2}{4R_0^2})H(0)^2+m^2
\ee
with
\bee
H^2(0) = R_0^2\left[2+\gfive \epsilon(h(-R_0)) + \epsilon(h(-R_0)) \gfive\right]\ .
\label{eq:1}
\ee

Since $[H^2(m),\gfive]=0$, $H^2$ can be written as a sum of two operators
each of which  acts only on one chiral sector, $H^2=H_+^2+H_-^2$.
In a sector with topological charge $Q$, $H^2(0)$ has, assuming the correctness
of the vanishing theorem, $|Q|$ zero modes
of chirality $\sigma=\sign(Q)$ and  $|Q|$ eigenmodes with eigenvalue $4 R_0^2$
in the opposite chirality.
The spectra of $H_+(0)^2$ and $ H_-(0)^2$
outside of eigenvalues at $\lambda = {0,4R_0^2}$
are identical. Changing the topological sector therefore amounts to transforming
a pair of eigenvalues with opposite chirality and eigenvalues $\lambda$
to eigenvalues 0 and $4 R_0^2$, i.e. a non-continuous change of the spectrum.
The rest of the spectrum will also change significantly.
This will become important in the discussion of the rate at which 
the topological sector changes.

We will construct the sign function $\epsilon(h(-R_0))$  by projecting out
the lowest $n_{\rm eig}$ eigenmodes $|\lambda\rangle$ of the kernel operator (with respect to modulus) and
using a Zolotarev approximation of order $n_z$ for the rest of the spectrum~\cite{Akhiezer1,Akhiezer2},
\bee
\epsilon(h)=\sum_{\lambda} \sign(\lambda) P_\lambda +
\sum_{i=1}^{n_z} h \frac{b_i}{h^2+c_i}  (1-\sum_{\lambda}P_\lambda),
\label{eq:5}
\ee
with $P_\lambda=|\lambda\rangle\langle\lambda|$ the projector on the eigenstate $|\lambda\rangle$.
We fix $n_z$, $b_i$ and $c_i$ such that the precision of the 
sign function between at least $\lambda_{n_{\rm eig}}$ and the maximal retained eigenvalue
of the kernel is better than some given value, e.g. $10^{-7}$.

The action of the sign function on a vector $\phi$ is computed by using a multi-mass \cite{ref:multimass}
 conjugate gradient (CG) to invert  $h^2+c_i$, after projecting out the
eigenmodes of the kernel operator from the vector. We will refer to this as the inner
CG as opposed to the outer CG used to invert $H^2(m)$.
We calculate the eigenmodes by the conjugate gradient method of
Ref.~\cite{Kalkreuter:1995mm}. We choose an accuracy for the step function and monitor
the norm of the vector $\epsilon(h)\phi$ during the CG iteration, stopping when either the
the observed accuracy of the step function is achieved, or when the inner CG
converges. (Of course, we wish to set the accuracy of the inner CG high enough, that the second
result never happens.)

 As proposed in Ref.~\cite{Cundy:2004pz} we
adjust the accuracy of the inner CG, $r_{\rm in}$, depending on the current residue
of the outer CG $r_{\rm out}$ according to
\bee
r_{\rm in}= {\rm min}(r_0 \frac{|b|}{r_{\rm out}},r_{\rm cut})
\label{eq:rcut}
\ee
with $b$ the  source vector of the inversion, $r_0$ the target residue
and $r_{\rm cut}$ a maximal inner residue.
Since we want to make sure that this does not introduce an additional inaccuracy 
we restart the CG after convergence with the inner residue fixed to $r_0$. With a good choice
for $r_{\rm cut}$ fewer inner CG steps will be needed in the first pass
and the restarted CG will converge in 1-2 steps. 

\subsection{The kernel operator}
The choice of the kernel Dirac operator used in the construction of
the overlap provides an opportunity of optimization. Most groups use
the Wilson operator because of the low computational effort
to apply it on a vector. One might also choose an operator which almost fulfills
 the Ginsparg-Wilson relation like the hyper-cubic actions proposed in
Refs.~\cite{Gattringer:2000qu,Hasenfratz:2000xz} and tested
 in \cite{Gattringer:2003qx}. This has the advantage that
the number of operations to `transform' this operator into the 
overlap operator is low, e.g. it takes a small number of 
inner CG steps. However, the application of the kernel operator alone
is much more expensive than for the Wilson action, involves more communication,
and the fermionic force is cumbersome to code.
We wish to find an optimum in between these two approaches. The application
of the operator should be as cheap as possible  with respect to computation and 
communication. But is also should approximate the Ginsparg-Wilson circle in 
some way.

The optimization has two parts: the choice of a free field action and the choice of
a gauge connection.

We use an action
similar to the one studied in Ref.~\cite{DeGrand:2000tf} which has nearest
and diagonal neighbor interactions. To be precise let us parameterize the 
associated massless free action by
\bee
S=\sum_{x,r} \bar \psi(x)\left[ \eta(r) + i \gamma_\mu \rho_\mu(r)\right] \psi(x+r)
\label{eq:fermionaction}
\ee
with  $r$ connecting nearest neighbors ($\vec r=\pm\hat\mu$;
$\eta=\eta_1$, $\rho_\mu = \rho_1$) and diagonal
neighbors ($\vec r=\pm\hat\mu \pm\hat\nu$, $\nu\ne\mu$;
$\eta=\eta_2$, $\rho_\mu= \rho_\nu = \rho_2$.
The constraint $\eta(r=0)=\eta_0 = -8\eta_1 -24 \eta_2$
enforces masslessness on the spectrum, and $-1 = 2 \rho_1 + 12\rho_2$ normalizes
the action to $-\bar \psi i \gamma_\mu \partial_\mu \psi$ in
the naive continuum limit. Thus there are three free parameters to choose.
These three parameters can be reduced to one by requiring that each of the couplings
of a fermion to its neighbors is a projector, proportional to $1 \pm \hat n \cdot \gamma$.
This is a familiar trick for Wilson action simulations. There, one gains almost
a factor of two in speed from the trick because  the multiplication of
the gauge connection times the spinor only needs to be done on two, not
four Dirac components, and because the ``gather'' of a spinor on a neighboring site
does not need all four Dirac components, only the linear combination of two
components which participates in the projection. For nearest neighbors,
a projector action corresponds to the constraint $\eta_1 = \rho_1$ (up
to signs) and for the diagonal neighbors, $\eta_2 = \sqrt{2}\rho_2$.
With this action, the gain in speed on a single-node computer
is only about 8-10 per cent, since the diagonal projection itself has an overhead.
Of course, on a parallel machine, projection halves the necessary internodal communication.

\begin{figure}[tb]
\begin{center}
\includegraphics[width=6cm,clip]{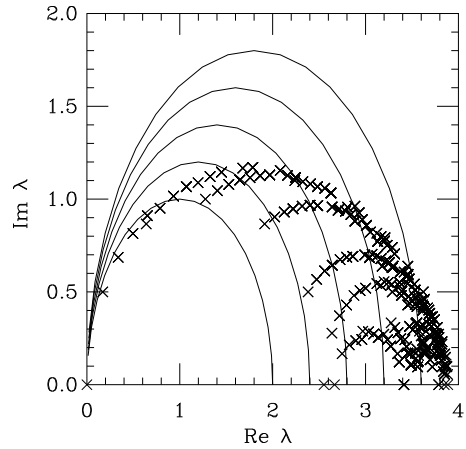}
\end{center}
\caption{Eigenmode spectrum of the kernel action used in this work on a finite lattice,
with superimposed circles (which show the spectrum of overlap actions).
	The five circles have radius 1.0, 1.2, 1.4, 1.6, 1.8.
}
\label{fig:planar}
\end{figure}

The action we use in the simulations presented in this paper uses  $\rho_1 = -1/6$ and $\rho_2 = -1/18$.
We also add a clover term with the tree-level clover coefficient appropriate to this action of 1.278.
For these parameters it seems optimal to set the radius of the Ginsparg-Wilson circle $R_0$ to 
1.2 since it is closest to the behavior of the kernel for the low-lying eigenmodes. 
We tested this value by timing a quenched approximation calculation of eigenmodes, for various
$R_0$ values. This was done
with Wilson action gauge fields, at coupling $\beta=5.9$, or a lattice spacing of about 0.13 fm.
One could vary $R_0$ by $\pm0.2$ without much effect. In Fig.~\ref{fig:planar}
we show the free field spectrum of the kernel Dirac operator we use together with several
Ginsparg-Wilson circles. Let us remark that contrary to the Wilson operator, 
which has a maximal eigenvalue
at 16, this operator has it at about 4. This alone reduces the condition number 
$\kappa=\lambda_{\rm max}/\lambda_{\rm min}$ of the
$h^2$, which controls the  convergence of the inner CG,  by a factor of about 16.

The other ingredient of our kernel action, the stout link as a gauge connection, is so important
that it deserves a separate discussion below, in Sec.~\ref{sec:fat}.

\subsection{Hybrid Monte Carlo}
We consider simulations with two flavors of degenerate-mass fermions. In a completely
standard approach, we add to the gauge action a set of momenta $\pi_\mu(n)$
(with total energy $\frac{1}{2}\sum_{\mu,n}\pi_\mu(n)^2$). To include the fermions,
we introduce their action using pseudo-fermions
\bee
S_f[U,\phi^\dagger,\phi] = \phi^\dagger H(m)^{-2} \phi
\ee
where the pseudo-fermion fields $\phi$ are a set of four-component spinors. At the beginning
of each trajectory we perform a heat-bath initialization of the fermion action by casting
a vector of four-component colored Gaussian random numbers $R$ and initialize $\phi = H R$.

We want to study the effect of additional pseudo-fermion fields as advocated in 
Refs.~\cite{Hasenbusch:2001ne,Hasenbusch:2002ai} on the performance of the HMC
algorithm. One therefore rewrites the fermion determinant of two degenerate
fermions as
\bee
\det H^2(m) =  \det H^2(m_{N_p})\prod_{i=1}^{N_p-1} \det \frac{H^2(m_i)}{H^2(m_{i+1})}
\label{eq:det}
\ee
with $m_1=m$ and $m_i<m_{i+1}$ with suitably chosen larger masses. Using pseudo-fermions $\phi_i$
to deal with the determinants, this leads as usual to the effective action

\bee
S_{\rm eff}[U,\phi^+_i,\phi_i]=S_g[U]+S_f[U,\phi^+_i,\phi_i]
\ee
with $S_g[U]$ the gauge action and  
\bee
S_f[U,\phi^+_i,\phi_i]=\phi^+_0H^{-2}(m_{N_{p}})\phi_0 +  
\sum_{j=1}^{N_{\rm p}-1} \phi^+_j\frac{H^{2}(m_{j+1})}{H^{2}(m_j)}\phi_j .
\ee

To calculate the fermion force one has to know the
variation of the fermion action. Since $[H^2,\gfive]=0$ one
can split the fermionic action into two contributions of definite
chirality. In the following we restrict ourselves to one pseudo-fermion
field. The generalization to multiple fields is trivial. For a chiral source
$\phi$ with chirality $\sigma=\pm 1$  we follow \cite{Gottlieb:1987mq} and \cite{Bode:1999dd} and
compute the simulation time derivative of the fermionic action
\bee
\frac{\rm d}{{\rm d} \tau} S_f[U,\phi^+,\phi]=
\phi^+\frac {\rm d}{{\rm d} \tau} H^{-2} \phi
=- 2 ( R_0^2-\frac{m^2}{4}) \sigma \psi^+ \frac{\rm d}{{\rm d} \tau} \epsilon(h)  \psi
\label{eq:6}
\ee
with $\psi= H^{-2}\phi$. This is only valid as long as no eigenvalue of $h(-R_0)$ is zero. 
Since an eigenvalue changing sign causes a step in the action, 
we get an additional contribution
\bee
\frac{\rm d}{{\rm d} \tau} S_f[U,\phi^+,\phi]\big |_{\rm disc}=\delta(\lambda) \Delta S_f[U] \dot \lambda \ . \label{eq:disc}
\ee

It is very difficult to approximate the derivative of the step function by the derivative
of its Zolotarev approximation, because near topological boundaries eigenmodes of $h$ will develop
arbitrarily small eigenvalues.
In order to compute the variation of the approximation to the 
sign function Eq.~(\ref{eq:5}), we have to know the variation of 
the projectors. This can be taken into account by using first
order perturbation theory \cite{Narayanan:2000qx},
\bee
\delta P_\lambda=\frac{1}{\lambda-h}(1-P_\lambda)\delta h P_\lambda+
P_\lambda \delta h^+\frac{1}{\lambda-h}(1-P_\lambda) .
\ee
The derivative of the approximation of the sign function used in  Eq.~(\ref{eq:6}) thus becomes
\bee
\psi^+\frac{\rm d}{{\rm d} \tau}\epsilon(h)  \psi 
= \psi^+ (1-P)\frac{\rm d}{{\rm d} \tau} \tilde \epsilon (h)(1-P) \psi
+
\sum_\lambda \big \{-
\chi_\lambda^+\dot{h}\tilde \rho_\lambda
-
\tilde \rho_\lambda^+\dot{h} \chi_\lambda
+
\epsilon(\lambda)(\rho_\lambda^+\dot{h}\tilde \rho_\lambda
+
\tilde \rho_\lambda^+\dot{h} \rho_\lambda) \big\}
\label{eq:force1}
\ee
with $P=\sum_\lambda P_\lambda$ and $\tilde \epsilon (h)$ the Zolotarev approximation to
the sign function. The $\chi$ and $\rho$ are defined as
\bee
\rho_\lambda=\frac{1}{\lambda-h}(1-P_\lambda)\psi\ ; \hspace{1cm}
\tilde \rho_\lambda=P_\lambda \psi\ ;\hspace{1cm}
\chi_\lambda = \frac{1}{\lambda-h} \epsilon(h) (1-P) \psi \ .
\ee
The derivative of the Zolotarev approximation can be simplified to
\bee
\frac{{\rm d}}{{\rm d}\tau} \sum_l h \frac{b_l}{h^2+c_l}
=\sum_l \frac{1}{h^2+c_l}\big[c_lb_l \dot h-b_lh\dot{h}h\big]\frac{1}{h^2+c_l} \ .
\ee
The  term in Eq.~(\ref{eq:disc}) comes from the derivative of the step function at
$\lambda=0$ where the height of the step in the fermionic action is $\Delta S_f$.
For this term we use now the Feynman-Hellmann theorem 
\bee
 \dot{\lambda}=\chi^+ \dot{h} \chi \label{eq:FH}
\ee
with $\chi$ the zero-mode of the kernel operator $h$. This force acts only if the
eigenvalue $\lambda $ of the  kernel operator $h(-R_0)$ changes 
sign. At these points the topology changes, because using Eqs.~(\ref{eq:hov0}) and (\ref{eq:sign})
\bee 
Q=\Tr H_{\rm ov} = \sum_\lambda \sign (\lambda) \ .
\label{eq:2}
\ee
To compute the height of this step in the action, we evolve the gauge fields
in simulation time onto the $\lambda=0$ surface. Then we compute $S_f$
on both sides of the surface, where we just change the sign of the value of the crossing
 eigenmode. This
is the only source of discontinuity. Anything else which might change across this
surface does so continuously and is thus captured by the equations of motion.
Then we use the prescription of Ref.~\cite{Fodor:2003bh} to alter the momentum $\pi$ depending on
whether the step in potential $\Delta S_f$ is high enough to reflect at the surface or whether
one can refract into the other topological sector. The new momenta then are
\bee
\Delta \pi = 
\begin{cases}
-N \; \langle N |\pi \rangle + N \; \sign  \langle N | \pi \rangle \;  \sqrt {\langle N | \pi \rangle^2-2 \Delta S_f} 
& \text{if  $\langle N | \pi \rangle^2>2 \Delta S_f $}\\
-2  N \langle N | \pi \rangle &  \text{if $\langle N | \pi \rangle^2\leq 2 \Delta S_f$}
\end{cases}
\label{eq:3}
\ee
where $\Delta S_f$ is the height of the discontinuity, $N$ the vector normal to the surface of zero eigenvalue
and $\pi$ the molecular dynamics momenta. The scalar product is defined by
\bee
\langle N | \pi \rangle = \sum_x \Tr (N^+ \pi) \ .
\ee
Using Eq.~(\ref{eq:FH}),
the normal vector $N$ is computed using the derivative of the kernel action with respect to $U$
\bee
N(x,\mu)= \left [ U(x,\mu) \langle \chi |\frac{\delta h}{\delta U(x,\mu)}|\chi \rangle
 \right]_{\rm TA} \ . 
\ee
with $|\chi \rangle$ the zero-mode of the kernel operator and we take the traceless anti-Hermitian 
part so that $\pi$ itself stays traceless and anti-Hermitian. 

A word on  the evaluation of $\rho_\lambda$: Since $\lambda$ is
an eigenvalue of $h$, the matrix $1/(\lambda-h)$ is singular. Although
$1/(\lambda-h)(1-P_\lambda)$ is well defined, in a numerical calculation
$\lambda$ and $|\lambda \rangle$ may not be sufficiently accurately known for the inversion to
be stable. In our
simulation we therefore compute $1/(\lambda+\delta_i-h)(1-P_\lambda) \psi$
for several $\delta_i$  and interpolate to $\delta=0$
afterward. 

In our simulation the kernel operator is constructed from stout links. Thus,
after calculating the variation of the kernel operator $\delta h$ with respect to the stout links
one has to apply the chain rule as described in Ref.~\cite{Morningstar:2003gk} to get
the variation with respect to the unsmeared link variable $U(x,\mu)$.

In order to maintain a  high acceptance rate, it is crucial to monitor whether a mode has crossed
the zero eigenvalue surface. Fortunately, functions of the gauge field variables,
 including eigenmodes of the kernel, evolve slowly in simulation time.
The identification of the modes before and after a molecular dynamics
time step is done by computing their scalar products. If the product of the vector at the 
beginning of the step with one at the end has an absolute value close to one, the 
two modes are identified. A technical problem occurs  if the modes become degenerate
during the time step since the wave functions can mix in this case. However,
enough similarity remains to identify the pair of the two modes which have mixed.
If we have identified that at least the eigenvalue of one mode   has changed  sign
we use the van Wijngaarden-Dekker-Brent method \cite{NumRec}
to find the first  point in simulation time where an eigenvalue vanishes.

To integrate the equations of motion we
 employ a Sexton-Weingarten  scheme \cite{Sexton:1992nu}
setting the time-scale  of the gauge updates
to  $1/n_{SW}$ of the scale for the fermionic updates $\delta \tau$. 
Using the following notation for the updates
\begin{eqnarray}
T_{\rm PG}(\tau) &: \pi &\rightarrow \pi - \imu \tau \delta_U S_g[U]\nonumber\\
T_{\rm PF}(\tau) &: \pi &\rightarrow \pi - \imu \tau \delta_U S_f[U]\\
T_{\rm U}(\tau)  &: U &\rightarrow e^{\imu\tau \pi} U\nonumber
\end{eqnarray}
the updating algorithm is composed of the following elementary steps:
\begin{itemize}
\item[1)]
Update momenta with fermion force: 
\bee
\pi \rightarrow T_{\rm PF}(\delta \tau/2)\pi
\ee
\item[2)]
Update gauge fields: Do $n_{SW}$ steps of the gauge update:
\bee
U\rightarrow\big[T_{\rm PG}(\delta \tau/(2n_{SW})) T_{\rm U}(\delta \tau/n_{SW})
T_{\rm PG}(\delta \tau/(2n_{SW}))\big]^{n_{SW}}
\ee
\item[3)]
Calculate eigenmodes of the kernel operator, decide whether an eigenmode
has changed sign. If this is the case go to the time when this has happened,
change the momenta according to Eq.~(\ref{eq:3}) and evolve until the end
of the time step.
\item[4)]
Update momenta with fermion force.
\bee
\pi \rightarrow T_{\rm PF}(\delta \tau/2)\pi
\ee
\end{itemize}

\section{Simulation Parameters}
We wanted to test the algorithm at parameters appropriate to simulations which could be used
to do physics. Quenched overlap simulations become increasingly difficult as the gauge
coupling is carried into the strong coupling regime, as the cost of the inner CG grows.
We also fear encountering the region of the Golterman-Shamir vanishing mobility edge\cite{Golterman:2003qe}.
Thus, the largest lattice spacing $a$ which we feel we can tolerate is at about
$a=1/(6T_c)$ or about 0.2 fm at a nominal $N_f=2$ deconfinement temperature of 150 MeV.
For our first round of tests, we do not need an accurate determination of $a$. Accordingly,
we equilibrated $6^4$ volumes at parameter values near their values for deconfinement.
The crossover from the confined phase will be very rounded, and the deconfinement phase will not
be very physical, since Polyakov loops in all directions will order, but we will still
have a rough idea of the lattice spacing. After the simulation parameters are optimized,
we can go to bigger spatial volumes.

We simulated  using the L\"uscher-Weisz gauge 
action \cite{Luscher:1984xn}. 
We approximated the tadpole improved coefficients of Ref.~\cite{Alford:1995hw}.
Explicitly the action reads
\bee
S[U]  =  \beta_1 \sum_{pl} \frac{1}{3} \; \mbox{Re~Tr} \; [ 1 - U_{pl} ]
\; + \;
\beta_2 \sum_{rt} \frac{1}{3} \; \mbox{Re~Tr} \; [ 1 - U_{rt} ]
 + 
\beta_3 \sum_{pg} \frac{1}{3} \; \mbox{Re~Tr} \; [ 1 - U_{pg} ] \; ,
\label{eq:sgauge}
\ee
\begin{equation}
\beta_2 \; = \; - \; \frac{ \beta_1}{ 20 \; u_0^2} \;
[ 1 + 0.4805 \, \alpha ]
\; \; , \; \;
\beta_3 \; = \; - \; \frac{ \beta_1}{u_0^2} \; 0.03325 \, \alpha
\; ,
\end{equation}
with $U_{pl}$ the plaquette, $U_{rt}$ the $1\times 2$ rectangle, and $U_{pg}$ the perimeter-6
``parallelogram'' loop, and
\begin{equation}
u_0 \; = \; \Big( \frac{1}{3} \mbox{Re~Tr} \langle U_{pl} \rangle
\Big)^{1/4} \; \; , \; \; \alpha \; = \; - \;
\frac{ \ln \Big(\frac{1}{3} \mbox{Re~Tr} \langle U_{pl} \rangle
\Big)}{3.06839} \; .
\end{equation}
After running on $6^4$ lattices using about 200 trajectories we measured $u_0=0.86$ for
several parameter sets, with little variation. We therefore decided
to set $u_0=0.86$ for all our production runs. Even though this is only a rough estimate
and not a self consistent determination of $u_0$, we expect this to capture
most of the effect of the tadpole improvement.

In nearly all runs we applied two levels of isotropic stout smearing and chose the
smearing parameter $\rho = 0.15$, see Sec.~\ref{sec:fat}.
This value was determined by  maximizing the average value of the 
plaquette constructed from the fat links on a number of quenched configurations.
This is the same strategy as used in
Ref.~\cite{Hasenfratz:2001hp} to determine the parameters of HYP smearing.
We believe that any value in the range  between 0.1 and 0.2 would work nearly as well.

\begin{figure}[t!hb]
\begin{center}
\includegraphics[width=5cm,clip,angle=-90]{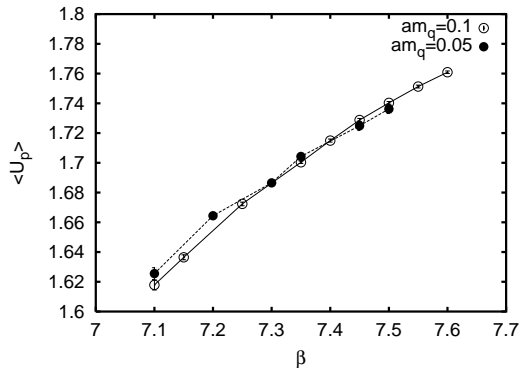}
\end{center}
\caption{
Plaquette  vs gauge coupling $\beta=10/g^2$ from dynamical overlap simulations on $6^4$ lattices
with quark masses $am=0.05$ and 0.1.
}
\label{fig:ploop}
\end{figure}

\begin{figure}[t!hb]
\begin{center}
\includegraphics[width=0.5\textwidth,clip,angle=-90]{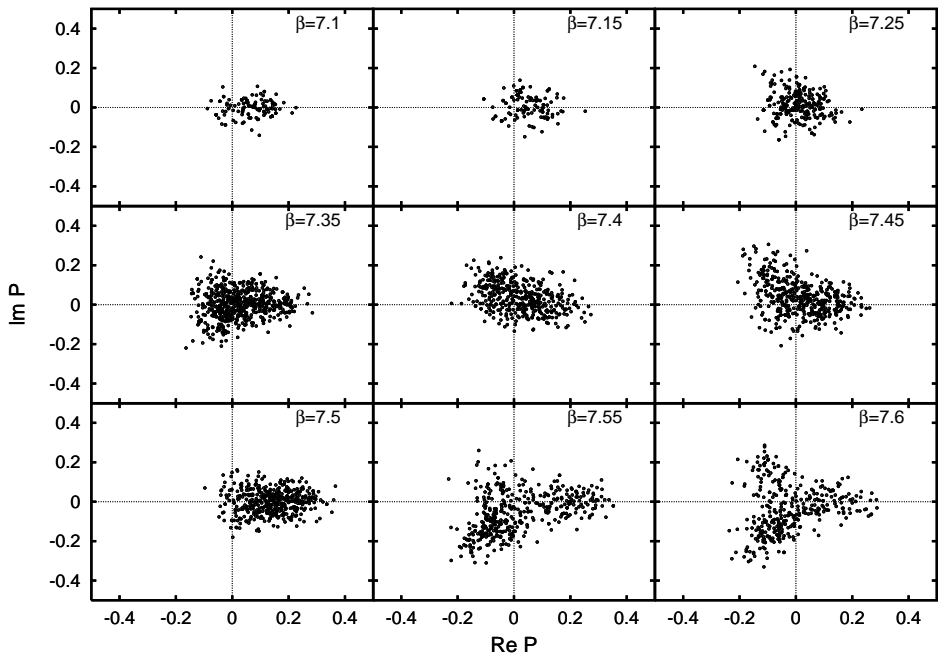}
\end{center}
\caption{
Scatter plot of the Polyakov loop 
with quark masses $am=0.1$ on the $6^4$ lattices for various values of $\beta$.
}
\label{fig:scatter}
\end{figure}

We performed runs at bare quark masses $am_q=0.05$, $0.1$ at $\beta$ values
such that we ran at both sides of the (pseudo) phase transition. A plot of the plaquette
vs. gauge coupling $\beta$ is shown in Fig. \ref{fig:ploop}. In order to illustrate
the change in the Polyakov loop as a function of $\beta$ we show a scatter plot at $am_q=0.1$ in Fig.~\ref{fig:scatter}.

\begin{figure}[t!hb]
\begin{center}
\includegraphics[width=5cm,clip,angle=-90]{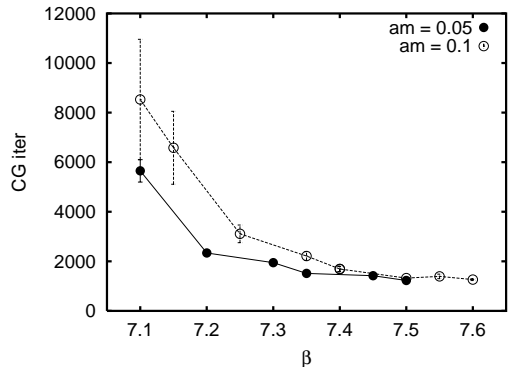}
\end{center}
\caption{Number of applications of the kernel operator on a vector per step
and pseudo-fermion field as a function of the gauge coupling. The runs
at $am_q=0.05$ where done with 2 pseudo-fermions, the runs at $am_q=0.1$ with three.
}
\label{fig:cg}
\end{figure}

We set the precision of the step function to be $10^{-7}$ and ran the 
CG for the fermionic action to a residue of $10^{-6}$.
For the HMC algorithm we chose to set the number of gauge update steps
$n_{SW}=12$ since this provided us with a gauge force of roughly the 
same size as the fermion force. Due to the high cost of the fermion inversions
the gauge  updates are still a negligible fraction of the cost the whole update.

A trajectory of molecular dynamics evolution had a typical time step of $\delta t = 1/20$ for
$am_q=0.05$ and $\delta t = 1/15$ for $am_q=0.1$.
At each step we computed $n_{\rm eig}=8$ eigenmodes of the kernel operator. 
The cost of these eigenmodes is $5\%$--$10\%$ overhead in the confined phase and
rises to as large as $50\%$ at high $\beta$ where the inversion is much cheaper.
One has to make sure that the gauge field moves between
two computations of the eigenvalues only such that an eigenvalue which
has changed sign is among the $n$ modes before and after the elementary step.
This along with the requirement that preferably not more than one eigenmode
changes sign per elementary step are the limiting requirements to the 
step size.

We ran simulations on our array of 31 old 800 Mhz P-III's and 12 new 3.2 Ghz P-IVE's,
all in single-processor mode. At $\beta=7.35$ and $am_q=0.1$ we
collect approximately 40 trajectories per day on the newer machines. Fig.~\ref{fig:cg} gives
the dependence on the gauge coupling
of the number of applications for the kernel operator on a vector per elementary step and
pseudo-fermion field. Our code is based on the
 MILC package\cite{ref:MILC},
with SSE routines\cite{ref:SSE} for all $SU(3)$ matrix multiplication.

We now describe explorations of various parameter choices.

First, there are some small tricks. We always begin the outer CG for the fermionic force with trial vectors
taken by interpolating solutions from the previous two time steps. We also extrapolate
the identified  eigenvectors of $h$ to begin the Conjugate Gradient calculation of new eigenmodes.
Careful tuning of the $r_{\rm cut}$ parameter (recall Eq. \ref{eq:rcut})
 can reduce the number of inner CG steps
(this from a $\beta=7.2$, $m=0.05$, $\delta \tau=1/20$ run) from about 35 to about 20.

\begin{figure}
\includegraphics[width=0.36\textwidth,angle=-90,clip]{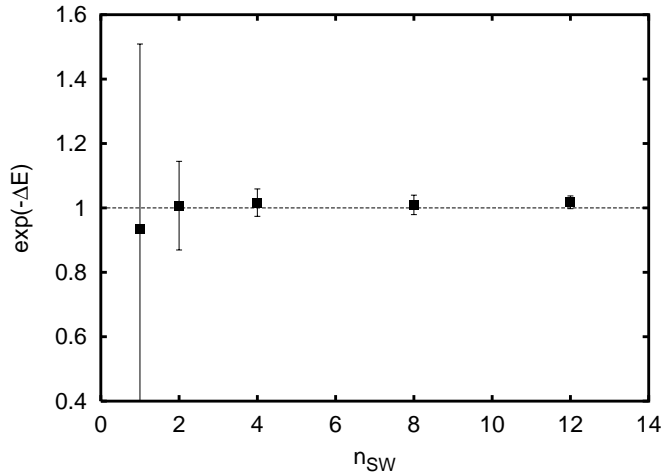}
\caption{\label{fig:deltahnsw}The average acceptance probability $e^{\Delta E}$ with
$\Delta E$ the violation in the microcanonical energy as a function of the number
of Sexton-Weingarten steps $n_{SW}$. We average over 19 trajectories which started
from the same equilibrium configuration at $\beta=7.4$ and $am_q=0.1$. The 
step-size is $\delta \tau=1/15$ and the trajectory length is 1 in all cases. The average
is always consistent with one as required by reversibility.}
\end{figure}

Fig.~\ref{fig:deltahnsw} illustrates the effect of Sexton-Weingarten updating. We plot the
average over  $e^{\Delta E}$ ($\Delta E$ being the energy violation after one molecular
dynamics trajectory of length 1) which governs the acceptance
probability.
It is from a run with $\beta=7.4$ and $am_q=0.1$ with time-step $\delta \tau =1/15$.
We started from the same equilibrium configuration and average over 19 trajectories.
We see a considerable reduction in energy non-conservation as $n_{SW}$ increases.
In this run, we can compare the fermion force per direction $\sqrt{ \sum_x \Tr F_{\mu}^+(x)F_{\mu}(x)}$
to the gauge force: the gauge force is about an order of magnitude larger. So setting
$n_{SW}\simeq 12$ reduces the gauge impulse to roughly the level of the fermion impulse.

This factor of 10 falls to about a factor of five in thin link simulations: the fermion force
doubles. This probably happens because thin link fermions do not decouple from high-momentum gluon modes.
A glance at a time history of $\Delta E$ shows
Sexton-Weingarten updating works less well; one would simply have to lower the molecular dynamics time step
$\delta \tau$ to reduce energy violation.

We typically run with either $N_p=2$ or 3 pseudo-fermion fields. As we are going to discuss
in Sec.~\ref{sec:pf}, this increases the number of changes of topology significantly whereas
it has  no impact on the step size. In order to check the implementation of the
additional pseudo-fermions we compared the plaquette for $N_p=1$, 2 and 3 fields.
They agreed within statistical errors.

\section{\label{sec:fat}Impact of fat links}
It has been long known that fat links can significantly decrease the computational
cost of applying the overlap operator on a vector \cite{DeGrand:2002vu}. They also
have a beneficial effect on the locality of the overlap operator \cite{Kovacs:2002nz}.
Recently  stout smearing  \cite{Morningstar:2003gk},  which is analytic in
the gauge fields and thus suitable for the HMC algorithm, has become available.
The stout link $\tilde U$ for one level of isotropic smearing is constructed from the gauge links by
\bee
\tilde U_\mu(x)= \exp(iQ_\mu) U_\mu(x) \ ; \
Q_\mu(x)=\left[ \frac{\rho}{i} V_\mu(x) U_\mu^+(x) \right]_{\rm TH}
\ee
with $V_\mu(x)$ the sum over all staples associated with the link and $[\cdots]_{TH}$ the traceless
Hermitian part of the matrix. The contribution of this smearing to the fermion force has
been worked out in Ref.~\cite{Morningstar:2003gk}.

A main effect of the smearing is to remove the majority of the  eigenvalues of $h(-R_0)$
close to 0. This has two effects. First, it implies that one has to compute fewer 
eigenvectors of the kernel operator to capture the ones within a certain distance to the origin. 
For the same number of inner eigenmodes the condition number of $h^2$ is therefore
smaller for the fat links than for thin links. This speeds up the inversion of the fermion
Dirac operator.
Furthermore, the computation of the fermion force is more stable since it
is mainly plagued by small eigenmodes which are not well separated. This also implies a
lower probability of two eigenmodes trying to change sign in the same HMC step, a problematic
part of this algorithm. It furthermore might  mean that the topology changes more rarely measured in
HMC time and therefore lead to a higher auto-correlation time.
However, it is not clear whether each of these attempts has the same probability of 
crossing over to the other topological sector or whether there is a correlation between
consecutive attempts. This can only be clarified by an actual simulation.

To get an  idea of the effect of this smearing, we started runs from an equilibrium configuration
at $\beta=7.2$ and mass $am_q=0.05$ with $\delta \tau =1/20$. We ran 16 trajectories with thin 
links, 101 with one level of stout smearing at $\rho=0.15$ and 591 with two levels of
smearing. The plaquette is $\langle \Tr U_p \rangle \approx 1.67$ and the 
plaquette made of stout links is 2.5  for one level of stout links and 2.8 
for two levels. In each of the runs we used  two pseudo-fermion fields. 
The results of these runs are summarized in Table~\ref{tab:stout}. We find that each level
of smearing reduces the cost per trajectory by roughly a factor of 3. The acceptance
rate is the same for thin links and one level of smearing, however, it improves
significantly for two levels. This is probably a result of the lower 
density of eigenmodes of the kernel action. 
The number of  reflections and 
refractions per trajectory decreases with more levels of smearing. It is more
than compensated for by the higher speed of the simulation.

\begin{table}
\begin{tabular}{c|c|c|c|c|c}
$N_{\rm smear}$& acc. rate & No. of traj & refl. & refr. & Mxv/traj. \\
\hline
0             & $69\%$    & 16          & 49    & 0     & $3.3\cdot 10^5$\\
1             & $66\%$    & 101         & 303   & 13    & $1.6\cdot 10^5$\\
2             & $93\%$    & 591         & 676   & 14    & $3.5\cdot 10^4$\\

\end{tabular}
\caption{\label{tab:stout}We give relevant quantities for the performance of the HMC algorithm
for different levels of stout smearing. The data is take at $\beta=7.2$ and $am_q=0.05$. We 
    list the number of stout steps, the acceptance rate, the number of trajectories and
    the number of reflections and refractions during these trajectories. The last column
    gives the average number of applications of the kernel operator on a vector per trajectory.}
\end{table}

\begin{figure}
\includegraphics[width=0.4\textwidth,angle=-90,clip]{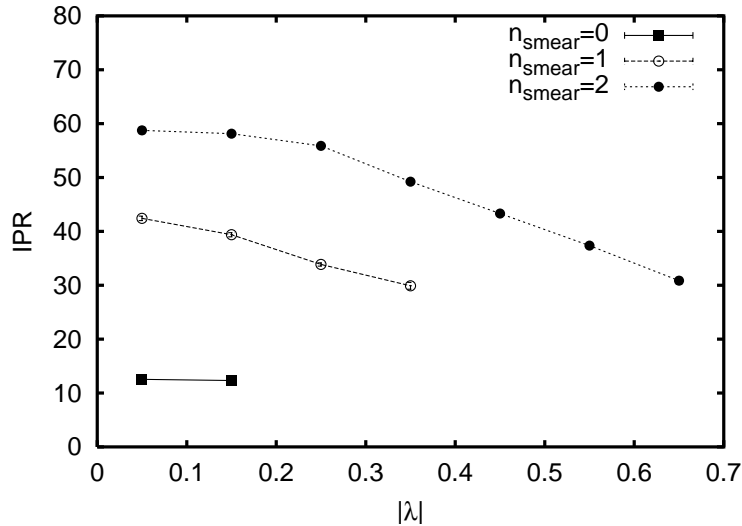}
\caption{\label{fig:ipr}The inverse participation ratio as a function of the modulus
of the eigenvalue of $h(-R_0)$. The data is from the 8 lowest eigenmodes only and
is binned in bins of size $0.1$. It is taken from a run  at $\beta=7.2$, $am=0.05$. The modes
of the Dirac operator constructed from thin links ($n_{smear}=0$) are only weakly
localized (small IPR). Each smearing step increases the localization of the modes. 
The spectrum of $h(-R_0)$ is denser near 0 for fewer smearing steps. Without 
smearing, we do not observe  eigenvalues larger than 0.18 among the 8 lowest modes.}
\end{figure}

Ref.~\cite{Golterman:2003qe} demonstrates
that insufficient localization of the low-lying modes of the kernel operator $h(-R_0)$ might have
implications on the locality of the overlap operator. We can get a rough measurement of the localization
of the modes through the inverse participation ratio ($V$ is the lattice volume)
\bee
I = V \sum_x [\psi^+(x)\psi(x)]^2.
\ee
If a mode is localized on a fraction $f$ of the lattice volume, we expect $I=1/f$.
Fig.~\ref{fig:ipr} shows a comparison of $I$ from a set of simulations at $\beta=7.2$, $am=0.05$,
with either one or two levels of $\rho=0.15$ stout smearing, or simulations with a thin link.
The inverse participation ratio (IPR) for the lowest eight modes has been binned and averaged.
We immediately see that eigenmodes of $h(-R_0)$ become progressively more localized
with increasing smearing. The thin link kernel eigenmodes are large on about ten per cent of
the lattice volume. This falls precipitously with smearing, to about two per cent for 
two smearing steps. We conclude that simulations with thin links at this set of parameter values
are not only expensive, they are dangerously close to delocalization.
(We are well aware, that we are comparing actions with different bare parameters
in the fermion sector.) Since the lattice spacing for different fermion actions (due
to different levels of smearing) might be different, we also computed the eigenmodes 
of the thin link $h(-R_0)$ operator on the configurations from the fat link data set.
We found essentially no difference to the IPR from the thin link data set.

We also notice that the range of eigenvalue of the lowest eight modes is considerably compressed
as smearing is removed. This means that the range of the Zolotarev  approximation of the
step function must be increased, with a consequent increase in its evaluation cost due
to the increased ill-conditioning of $h^2+c_i$.

\section{Changing the topological sector\label{sec:pf}}
When running with only one pseudo-fermion field we found it very difficult
to change topological sector, i.e. the algorithm reflected most of the time
at the boundary.
We observed that the height of the step was typically ${\cal O}(200)$ compared
to $|\langle N,\pi\rangle|^2$ which is ${\cal O}(1)$. This is the case independent
of the direction of the change in topology, e.g. whether one changes from
$Q=0$ to $Q=1$ or vice versa. It only occurs rarely that $\Delta S$ is
such that a change in topology is possible and one gets long sequences without
a change in topology indicating a high auto-correlation time. 

Why is this so? The $\phi$ field in the HMC algorithm with one pseudo-fermion field
is generated with
the distribution $\exp (-\phi^+ H^{-2}\phi)$. This is done by generating
a Gaussian random vector $R$ and then multiplying it by $H$. 
The expression which gives the contribution of the fermionic determinant
during the evolution is thus given by
 $\exp(-|H_1^{-1}H_0R|^2)$ with $H_0$ the Hermitian Dirac operator at the beginning of the
trajectory and $H_1$ the 'current' Hermitian Dirac operator. In Ref.~\cite{Hasenfratz:2002ym}
it was found that this is a good estimator for the change in the fermionic
determinant only if the eigenvalues of $H_0$ and $H_1$ are very similar. 
This is likely the case as long as one stays in the same topological sector.
However, the eigenvalue spectrum changes significantly when the topological
sector changes, e.g. two eigenvalues of $H$,  $\pm \lambda$, are transformed to
$\{m,2 R_0\}$ or vice versa.
To be precise, the change in the determinants is given by
\bee
\frac{\det H_1^+H_1}{\det H_0^+H_0}\propto\int dRdR^+e^{-R^+R}  e^{-R^+\Omega R} \ .
\ee
with $\Omega= H_0^+H^{-2}H_0-1$.
The exponential function is bounded by zero from below. So large
fluctuations in  $f[R]=\exp(-R^+\Omega R)$
mean lots of events with $f[R]\approx 0$ and only very few ones with $f[R]$
large. For $R^+\Omega R$ this means a large number 
of (very) large values and only very few small ones. Small fluctuations,
however, mean that $R^+\Omega R$ has values in a small interval around some
constant given by the determinant ratio.

\begin{figure}
\includegraphics[width=8cm,clip]{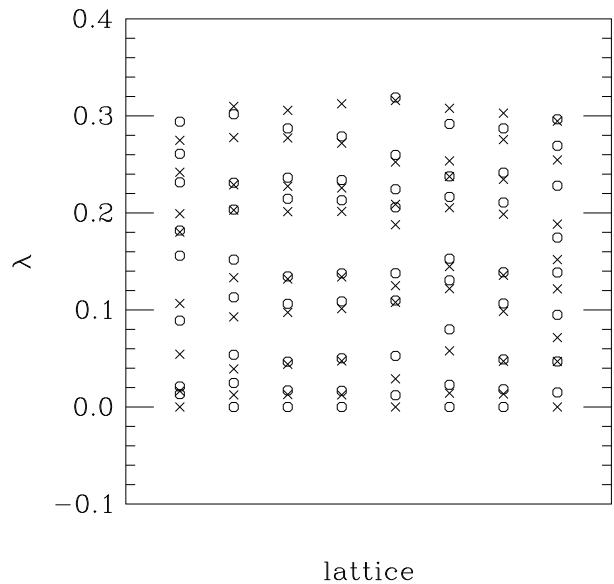}
\caption{\label{fig:change}Low-lying spectrum of $H(0)^2$ for a set of $\beta=7.3$, $m=0.1$ lattices
on either side of a topology-changing boundary. Octagons and crosses show the spectrum on either side.}
\end{figure}
Thus, while $H_1$ is in the same chiral sector as $H_0$, the
change in the fermionic action  is small. However, when we cross
to the other chiral sector, the large fluctuations in the estimator
are most likely such that $\Delta S_f$ is a large number, which
means that the trajectory will reflect from the boundary and stay
in the same topological sector. It is therefore pivotal to
employ a better estimator for the change in the determinant.

As an illustration of the change in the spectrum, we have computed a set of low-lying eigenmodes
of $H(0)^2$ on either side of a topology change. Specifically, we move onto the critical
surface
and compute the spectrum of $H(0)^2$ twice, once with the eigenvalue of the lowest eigenmode set
to its value just before the crossing, and then a second time with its eigenvalue's sign flipped.
The parameters are $\beta=7.3$, $m=0.1$. Fig. \ref{fig:change} shows that not only does a zero mode
appear or disappear, but all the low lying eigenvalues of $H(0)^2$ change discontinuously.
This is to be expected: eigenmodes of Hermitian matrices repel, and the 
appearance of a zero mode (for example)
must push nearby modes away.

To reduce fluctuations, we decided to use the method proposed in 
Refs.~\cite{Hasenbusch:2001ne,Hasenbusch:2002ai},
which consists of rewriting the fermion determinant as
in Eq.~(\ref{eq:det}). In this method, only determinant ratios are evaluated
using pseudo-fermions for the light quark masses. The change in the spectra
while changing topological sector of the ratio $H(m)/H(m')$ can be expected to be 
less dramatic than the change of the spectrum of  $H(m)$. Only the 
determinant of $H(m_N)$  is evaluated directly.
However, for a large mass $m_N$ the spectrum of $H^2$ is confined to a smaller 
region between $m_N^2$ and  $4 R_0^2$
and the change in the spectrum therefore less drastic than for a smaller mass.
In order to distribute the contribution to the fermionic action equally 
among the different pseudo-fermion fields we choose the $(N-1)$ auxiliary masses 
for a sea quark mass of $m_1$ to be
\bee
m_i= m_1^{(N-i+1)/N} \ .
\ee

\begin{figure}
\includegraphics[width=4cm,clip,angle=-90]{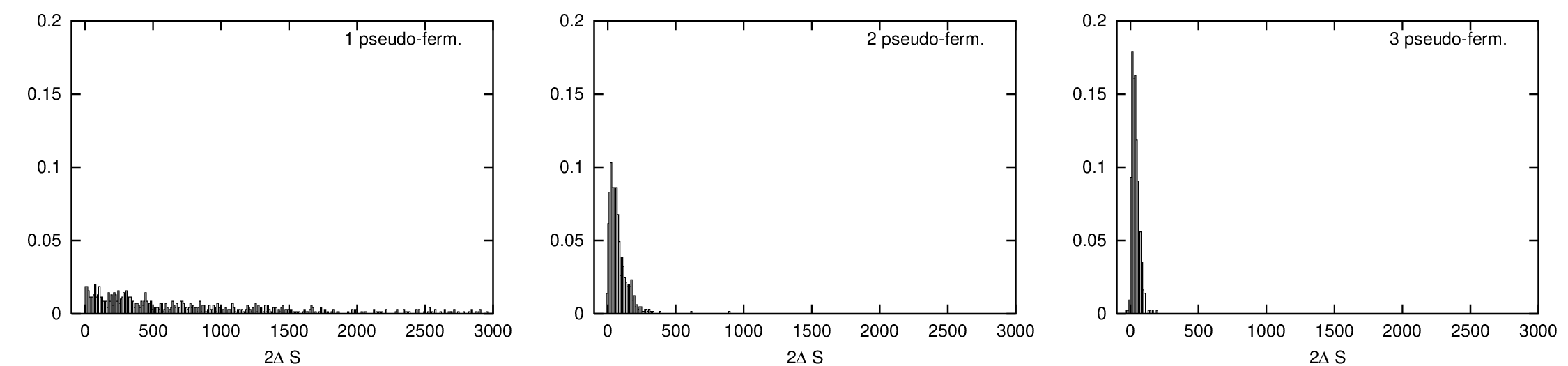}

\includegraphics[width=4cm,clip,angle=-90]{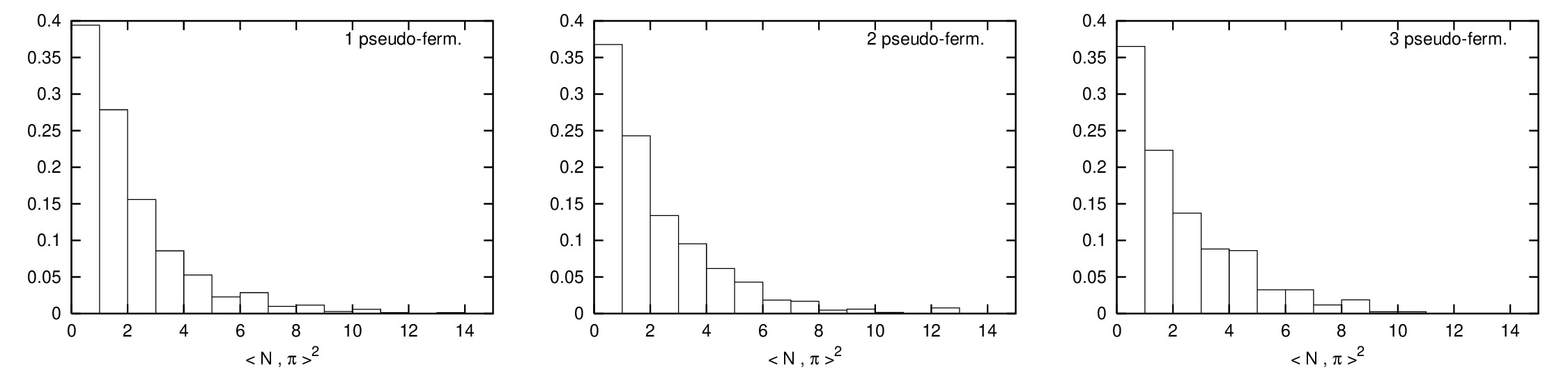}
\caption{\label{fig:deltas}Distribution of the  two components which determine 
whether a trajectory reflects or refracts at the boundary between two
topological sectors. Refraction takes place if $2\Delta S<|\langle N,H \rangle|^2$.
The data is taken at $\beta=7.2$ and $am_q=0.05$. For the 1 pseudo-fermion distribution
of $2 \Delta S$ we do not show about $5\%$ of the data which is above 3000.}
\end{figure}

In Fig.~\ref{fig:deltas} we show the distribution of $2\Delta S_f$ and
$|\langle N,\pi\rangle|^2$. The topological sector changes if 
$2\Delta S_f<|\langle N,\pi \rangle|^2$.
The data is taken from a run at $\beta=7.2$ and $am_q=0.05$ and $\delta \tau = 1/20$.
We see that for the standard HMC with one pseudo-fermion field the distribution
of  $2\Delta S_f$ is  spread out up to values of several thousand whereas the
distribution of $|\langle N,\pi \rangle|^2$ only spreads up to $\approx 10$.
(The distribution of  $|\langle N,\pi \rangle|^2$ is
independent of the number of pseudo-fermion fields.) 
The spread of $2\Delta S_f$ shrinks considerably with more pseudo-fermion
fields.

In these runs we observed roughly $1.5$ attempts per trajectory to cross over to another 
topological sector independent of the number of pseudo-fermion fields.
(For a summary of the results see Table~\ref{tab:hb}.)
The additional fields have a significant impact on the rate of topology 
changes. Whereas a change only occurs once in roughly 200 trajectories for
one pseudo-fermion field, it occurs once per 35 trajectories for three
pseudo-fermions. However, the additional fields seem to have no impact
on the acceptance rate. The additional cost of the auxiliary fields is thus
only justified by the higher tunneling rate and does not pay back 
in a possibility to decrease the step size as in the case of simulations
with Wilson fermions. It appears from Table \ref{tab:hb} that a single additional set of pseudofermions
is sufficient.

\begin{table}
\begin{tabular}{c|c|c|c|c|c}
No. of PF\ &\ trajectories \  &\  attempts/traj.\ &\  refr. prob.  \   &\  MatVec / traj.  \    &\ acc. rate\ \\
\hline
1  &602      & 1.3(1)      & $0.5(3) \%$        & $2.3(1)\cdot 10^5$  & $90\%$\\
2  &591      & 1.2(1)      & $2.0(6) \%$        & $3.4(1)\cdot 10^5$  & $93\%$\\
3  &290      & 1.6(2)      & $3(1) \%$        & $5.5(4)\cdot 10^5$  & $89\%$\\
\end{tabular}
\caption{The effect of additional pseudo-fermion fields. We list the
the number of trajectories we analyzed, the
attempts to change topology (reflections plus refractions)  per trajectory,
the percentage of refractions and the number of applications of $h(-R_0)$
on a vector per trajectory. In the last column we give the acceptance
rate.  The data is taken at $\beta=7.2$ and $am=0.05$.\label{tab:hb}
}
\end{table}

\begin{figure}
\includegraphics[width=4cm,clip,angle=-90]{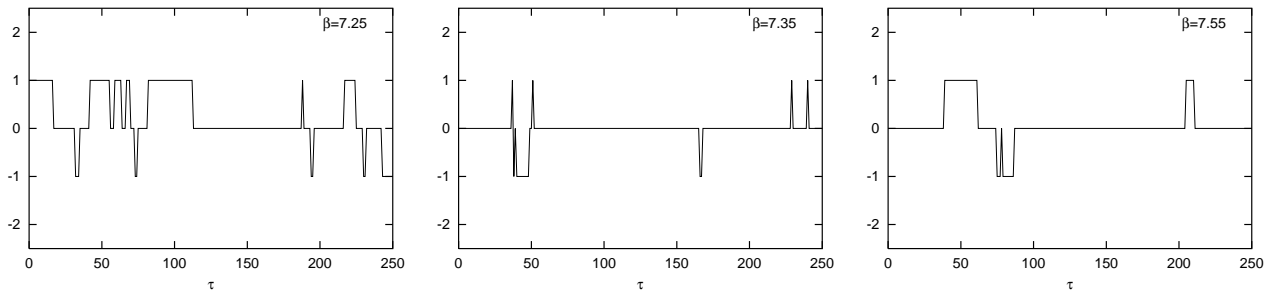}
\caption{\label{fig:topo}The history of the topological charge in simulation time for
$am_q=0.1$ at three values of $\beta$. $\beta=7.25$ is probably below the phase transition
whereas $\beta=7.55$ is above. The runs where done with three pseudo-fermion fields.}
\end{figure}

In Fig.~\ref{fig:topo} we show the time history of the topological charge
as a function of simulation time for three values of $\beta$ and $am_q=0.1$.
The runs were done with three pseudo-fermion fields. Below the phase 
transition we observe a lot of changes in topology and $|Q|$ up to 2. This
is as expected since chiral symmetry is expected to be broken and the lattice
spacing is large. With larger $\beta$ the lattice spacing gets smaller,
the temperature gets larger and we therefore expect fewer instantons and 
a smaller topological charge. This is confirmed by our simulations.

\section{Conclusions}
Simulations with two flavors of light dynamical overlap fermions with lattice spacing near
$a \simeq 1/(6T_c)$ and $am_q=0.05$ ($m\simeq 50$ MeV?) and small volume (a few fm${}^4$)
are certainly feasible on arrays of contemporary (2004) work stations. The crucial
ingredient needed to do such simulations is a fermion
action with a fat link gauge connection.

To be definite, we summarize our action: The gauge action is given by Eq.~(\ref{eq:sgauge})
with $u_0=0.86$. The kernel action is that of Eq.~\ref{eq:fermionaction}
with $\rho_1=-1/6$, $\rho_2=-1/18$, a clover term with coefficient $C_{SW}=1.278$,
 and the gauge connection is a  two-level stout link with smearing parameter
$\rho=0.15$. We use it in the overlap with (negative) mass shift $R_0=1.2$.

We are quite aware, that we have not completely characterized how much more efficient
fat link overlap actions are, than the conventional thin link ones. Table~\ref{tab:stout}
suggests a factor of 10 at equal $\delta \tau$ and a (probably a factor of  2)
smaller $\delta \tau$
is needed for equivalent acceptance rate. Suffice it to say,
thin link overlap is so expensive that we could not do anything useful with it with our
 available resources.
Observations such as Fig.~\ref{fig:ipr}  suggest that thin link overlap might
require a smaller lattice spacing than $a \simeq 1/(6T_c)$ to avoid delocalized kernel
eigenmodes.

The use of additional pseudo-fermion fields as suggested by Hasenbusch
is very beneficial to the rate at which the topological sector changes.
However, at our parameter values  it seems to have no
impact on the HMC acceptance rate.

Conventional large-lattice simulations are of course still not
possible with these actions with small computer 
resources, but there are physics issues whose solution is much more sensitive
to exact chiral symmetry than to large volume. We hope to address them in the near future.

\section*{Acknowledgments}

This work was supported by the US Department of Energy.
We are grateful to Z.~Fodor and A.~Hasenfratz for conversations and correspondence.
Simulations were performed on the University of Colorado Beowulf cluster. We are
grateful to D. Johnson for its construction and maintenance,
and to D. Holmgren for advice about its design.

\end{document}